\documentclass[11pt]{article}
\usepackage{moriond,epsfig,amssymb,graphicx,amsmath,hyperref}
\newcommand{\oper}{\mathcal{O}}
\newcommand{\CL}{\,\hbox{\rm CL}}

\def\Lag{\mathcal{L}}

\def\be{\begin{equation}}
\def\ee{\end{equation}}
\def\bea{\begin{eqnarray}}
\def\eea{\end{eqnarray}}

\begin{document}
\vspace*{4cm}
\title{ElectroWeak precision data: the minimal set of parameters}

\author{GUIDO MARANDELLA}

\address{Department of Physics, University of California, Davis, CA 95616, USA}

\maketitle\abstracts{We show how precision electroweak data, assuming CP conservation and flavor universality, despite being sensitive to 20 dimension 6 operators, constrain severely only 9 combinations of them. We define a set of 7 oblique parameters $\hat S, \hat T, \hat U, V, X,W,Y$ which fully describe corrections to the leptonic observables. We add two additional parameters $\delta \varepsilon_q, \delta C_q$ to take into account the most precise hadronic observables. Another parameter $\delta \varepsilon_b$ is considered for the bottom quark if flavor universality is not assumed for the third generation of quarks. We show that the approximation is extremely satisfactory testing it on a set of random new physics models and on some famous $Z'$ gauge bosons.}

\section{Introduction}

To make the Standard Model (SM) a natural low energy effective field theory some new physics is expected at the TeV scale. Such new physics has to cut-off the quadratic divegences which plague the SM, destabilizing the Higgs boson potential. A lot of possibilities have been explored in the last two decades.  Some of them try to enhance the symmetries of the SM (like supersymmetry), others try to make the Higgs a composite particle of the size of an inverse TeV (like technicolor and its 5-dimensional duals), others want to interpret the Higgs as a pseudo-Goldstone boson of some approximate global symmetry.
The astonishing success of the SM in the precision experiments performed in nineties mainly at LEP and SLD has pushed most of these models in rather innatural corners of their parameter space. 

Thus it is important to analyze in a complete, but possibily minimal way, what such precision experiments say. Global fits to all measured observables are of course a correct approach, but they don't give much intuition into the true origin of the strongest constraints. In fact, the number of severely constrained parameters is typically smaller then the total number of measured quantities.

Our goal is to define a minimal number of parameters which catch most of the effects on the observables measured so far. In this way a necessary and sufficient criterion of compatibility with the SM of a given new physics model can be defined. Possible large corrections to combinations of parameters which are accidentally poorly constrained are automatically taken care of if the parameters are properly defined.

In Sec. \ref{sec:operators} we define the full set of the 20 dimension 6 operators $\mathcal{O}_i$'s which affect precision observables. In Sec. \ref{sec:parameters} we show how only about half of them is severely constrained, and give an operative definition of $\hat S, \hat T, \hat U, V, X,W,Y,\delta C_q,\delta \epsilon_q$ (and possibly $\delta \epsilon_b$ for the third quark generation) . These are the only severely constrained paramaters. In Sec. \ref{sec:good} we verify quantitatively the goodness of our approximation. To be concrete, in Sec. \ref{sec:Zprimes} we apply our analysis to the case of $Z'$ gauge bosons. In Sec. \ref{sec:conclusions} we draw our conclusions.

\section{The operators affecting EW precision observables}
\label{sec:operators}

The set of precision experiments we refer to are $e^+ e^-$ collisions performed in the nineties at LEP and SLD \cite{:2004qh,Abbiendi:2003dh}. The experimental activity can be logically split into two parts. Collisions performed at the $Z$ resonance (LEP1 and SLD) and measurements above the $Z$ pole, up to energies $\sqrt{s} = 209 $ GeV (LEP2).

\begin{figure}
$$\includegraphics[width=0.5\textwidth]{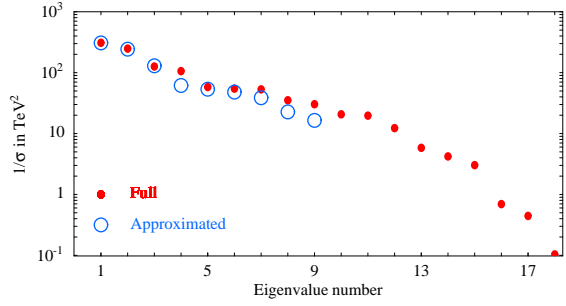}$$
\caption{\em \label{fig:sigmas} The red dots are the ordered eigenvalues of the full error matrix, that describe the sensitivity of present data (upper dots correspond to more precise combinations).
Precision data significantly constrain only about 10 new-physics effects.
The blue circles show the same eigenvalues recomputed making our simplifying approximation.}
\end{figure}

Effects of heavy new physics can be parametrized through a set of dimensions 6 operators $\oper_i$ added to the SM lagrangian:
\begin{equation} \label{eq:lag}
\mathcal{L} = \mathcal{L}_{\rm SM} + \sum_i c_i \oper_i
\end{equation}
To set our notation, we define the Higgs and fermion currents as
\begin{equation}
J_{\mu H} =  H^\dagger i \mathcal{D}_\mu H,\qquad
J_{\mu H}^a = H^\dagger  \tau^a i\mathcal{D}_\mu H,\qquad
J_{\mu F} = \sum \bar{F} \gamma_\mu F,\qquad
J_{\mu D}^a = \sum \bar{D}\gamma_\mu \tau^a D
\end{equation}
where $F = \{E,L,Q,U,D\}$ and $D=\{L,Q\}$.
Assuming CP conservation and an $U(3)^6$ flavor symmetry,  the relevant set of operators consists into \cite{Han:2004az}

\begin{itemize}
\item 7 operators involving one fermion and one Higgs current (vertex operators): $\mathcal{O}_{HF}=   J_{\mu H} J_{\mu F}+ \hbox{h.c.} $ with $F=\{L,E,Q,U,D\}$, and $\mathcal{O}'_{HD} =   J_{\mu H}^a J^a_{\mu D}+ \hbox{h.c.}$, where $D=\{L,Q\}$.
\item 11 4 fermion operators $\mathcal{O}_{FF'} = J_{\mu F} J_{\mu F'} / (1+\delta_{FF'})$ and $\mathcal{O}'_{DD'}  = J^a_{\mu D} J^a_{\mu D'}/(1+\delta_{DD'})$ involving at least one charged leptons current
\item 2 oblique operators, i.e. not involving fermions. They are $\mathcal{O}_{WB} = \left( H^\dagger \tau^a H \right)\, W^a_{\mu \nu} B^{\mu \nu}, \;\; \mathcal{O}_{HH} = |J_{\mu H}|^2$.
\end{itemize}

This makes a total of 20 parameters. Observables at and below the $Z-$ resonance are sensitive to only 10 of them \cite{Barbieri:1999tm}. They are all the 7 fermion-Higgs operators (vertex corrections), $\mathcal{O}'_{LL}$ (corrections to the muon decay which sets the Fermi constant), and the two oblique operators $\oper_{WB}, \oper_{HH}$. Going above the $Z$ pole, i.e. considering the LEP2 data, one becomes sensitive to 10 more operators: the remaining 10 4-fermi operators.

At this point, given a certain model of New Physics, one can calculate the corrections to the 20 parameters and perform a global fit \cite{Han:2004az}. However, one might wonder if all these 20 parameters are really constrained. For example, it has been recently pointed out \cite{Grojean:2006nn} that two combinations of these 20 parameters are almost unconstrained. For example, after calculating a global $\chi^2$ in terms of the coefficients of the 20 parameters, one can look to the eigenvalues of the error matrix. This automatically identifies all correlations of theoretical, experimental and accidental nature. The eigenvalues are shown in Fig \ref{fig:sigmas}. It is clear that only about 10 combinations of these operators are important, and that a few constraints often dominate the fit. 

It is interesting, at this point, to try to define the mostly constrained parameters which appear in Fig. \ref{fig:sigmas}. This will be done in the next section.

\section{The minimal set of parameters}
\label{sec:parameters}

In order to identify the mostly constrained parameters it is convenient to use the equations of motion for the gauge bosons. In this way one obtains an equivalent set of operators, but in the new basis it will be more transparent what are the important parameters. At leading order, the equations of motions are
\begin{align} \label{sys:eqs}
\partial^\nu B_{\nu \mu} + \frac{M_W^2}{g^2} g' (g' B_\mu - g W^3_\mu) + g' \sum_F Y_F J^f_{\mu} & =  0 +\ldots \\
\partial^\nu W^3_{\nu \mu} + \frac{M_W^2}{g^2} g (g W^3_\mu - g' B_\mu) + g \sum_f T_3 J^f_\mu & =  0 +\ldots \\
\partial^\nu W^{\pm}_{\nu \mu} + M_W^2 W^+_\mu + \frac{g}{\sqrt{2}} \sum_F J^{\pm}_{\mu F} & =  0 + \ldots
\end{align}
where we neglected on the r.h.s.\ operators that are poorly
measured. We now  solve the equations of motion in terms of
\begin{equation}
J^{e_R}_\mu \equiv J_{\mu E}= \bar e_R \gamma_\mu e_R ,\qquad
J^{e_L}_\mu\equiv \bar e_L \gamma_\mu e_L ,\qquad
J^{+}_{L\mu} \equiv \bar e_L \gamma_\mu \nu_L
\end{equation}
and plug the result into the Lagrangian generated by the new
physics. In this way the charged lepton currents disappear from the Lagrangian. Vertex corrections and 4-fermi operators involving only leptons are now recast into oblique corrections. The basis of 20 parameters defined in the previous section is now mapped in the following basis \footnote{For detailed formulas about the connection between the two basis see Appendix A of \cite{Cacciapaglia:2006pk}}.
\begin{itemize}
\item 7 oblique parameters  $\hat{S},\hat{T},\hat{U},V,X,W,Y$ . They can be expressed in terms of the gauge bosons self-energies $\Pi_{ij}(p^2)$ as
\begin{equation}
\begin{array}{c}\displaystyle
\hat S = \frac{g}{g'} \Pi_{30}'\,, \quad \hat T = \frac{\Pi_{33} - \Pi_{WW}}{M_W^2}\,, \quad W = \frac{M_W^2}{2} \; \Pi''_{33}\,, \quad Y = \frac{M_W^2}{2} \; \Pi''_{00}\,,\\ \displaystyle
\hat U =\Pi'_{WW} -  \Pi'_{33} \,, \qquad V = \frac{M_W^2}{2} \;
(\Pi''_{33} - \Pi''_{WW})\,, \qquad X = \frac{M_W^2}{2} \;
\Pi''_{30}\,,
\end{array}
\end{equation}
\item Vertex corrections for fermions other than charged letpons. They can be parametrized as
\begin{equation}\label{eq:vert}
\Lag_{\rm couplings} = \sum_f (\bar f \gamma^\mu f) \left[ e\, A_\mu\, \frac{C^\gamma_f}{M_W^2}\, p^2
+ \sqrt{g^2+{g'}^2}\, Z_\mu \left( \frac{C^Z_f}{M_W^2}\, (p^2 -
M_Z^2) + \delta g_{f} \right) \right]\,, 
\end{equation}
where $f={u_L,d_L, u_R,d_R,\nu_L}$. The $\delta g$'s are corrections to
on-shell $Z$ couplings, tested by measurements at the $Z$ pole. The
$C^\gamma$ and $C^Z$ are equivalent to 4-fermion contributions to
$e^+ e^- \rightarrow q \bar q$. $\delta g_{L\nu}$ can be written as linear combination of oblique parameters \cite{Cacciapaglia:2006pk}. Thus it is properly included in the fit considering the full set of 7 oblique parameters. Furthermore, only 11 of the 12 are independent, since the relation $(C^\gamma_{dL}-C^\gamma_{uL}) = \cos^2 \theta_W (C^Z_{d L}-C^Z_{u L}) + X/\tan \theta_W$ holds.
\end{itemize}

At this point we are left with 18 independent parameters. The 7 oblique ones, 4 $\delta g$'s for the
quarks, 4 quark $C^Z_q$'s and 3 independent $C^\gamma_q$'s. The total is 18 parameters: 2 less then the 20 we started from. This is because we neglected the poorly constrained operators which affect only trilinear couplings among vectors \cite{Grojean:2006nn}.

In the new basis it is very transparent to understand which are the most important parameters, All the corrections involving only leptons are included in the 7 oblique parameters $\hat{S},\hat{T},\hat{U},V,X,W,Y$. These are 
\begin{equation}
\alpha_{\rm em},  ~\Gamma(\mu), ~
M_Z,  ~M_W, ~ \Gamma(Z\to \ell\bar\ell),~ A^\ell_{FB}, ~A^\ell_{LR},
~ A_{\rm pol}^\tau, ~\sigma_{\rm LEP2}(e\bar e\to \ell\bar\ell), ~ee\to
ee
\end{equation}
They are the most precisely measured observables, tested at the {\it per mille} level. In most of the cases they are a sufficient set of  parameters. However this approximation would fail if, for some reason, new physics is leptophobic, i.e. if quarks are affected much more strongly then letpons. In this case one has to add some additional parameter to include the hadronic observables, i.e.
\begin{equation}
\Gamma(Z\to q\bar q),~
A^b_{FB}, ~ A^b_{LR},~ A^c_{LR},~A^c_{FB},
\sigma_{\rm LEP2}(e\bar{e}\to q\bar{q}),~ Q_W\
\end{equation}

It turns out that the most precise measurements are, at the $Z-$ pole, the hadronic branching fraction of the $Z$ and, above the pole, the total cross section $e \bar e \to $ hadrons. One can show \cite{Cacciapaglia:2006pk} that corrections to these two observables depend mainly on the two following combinations, which are the 2 additional parameters we include in our fit. 
\begin{equation}
\delta \varepsilon_q  =  \delta g_{uL} - \delta g_{dL}\,, \;\;\;\;\;
\delta C_q   = C^Z_{uL} - C^Z_{dL}\,.
\end{equation}

Furthermore in many models of electroweak symmetry breaking 
the third generation of quarks is special due to the heavyness of the top quark, and it is 
differently affected by new physics. For this reason, one can relax the flavor universality for 
the bottom quark, and deal with it separately. This is also necessary since the bottom final 
state is well measured. The most relevant parameter is the left-handed bottom quark coupling $\delta g_{bL}$ measured at LEP1. Thus we define $ \delta \varepsilon_b$ as
\begin{equation}
\delta g_{bL} = -\frac{1}{2} \delta \varepsilon_b \;.
\end{equation}

Summarizing, we defined a set of $7+2+1=10$ parameters. The 7 oblique parameters $\hat{S},\hat{T},\hat{U},V,X,W,Y$ fully describe the effects on all the leptonic observables, which are the best measured ones. 2 additional parameters $\delta \varepsilon_q, \delta C_q$ describe the most constrained quantities for the hadronic observables, and 1 additional parmeter $\delta \varepsilon_b$ describes deviations in the left-handed bottom quark coupling to the $Z$ if flavour universality is not assumed for the third generation of quarks. If Ref. \cite{} best fit, errors and correlations for all the 10 parameters is given.
In the case on universal models, only 4 parameters are non-vanishing: $\hat{S},\hat{T},W,Y$ \cite{Barbieri:2004qk}.

\section{Goodness of the approximation}
\label{sec:good}

%\begin{figure}[tb]
%\begin{center}$$\hspace{-0.06\textwidth}
%\includegraphics[width=0.4\textwidth]{his1}
%\hspace{-1cm}
%\includegraphics[width=0.4\textwidth]{his2}
%\hspace{-1cm}
%\includegraphics[width=0.4\textwidth]{his3}$$
%%\includegraphics[width=1.05\textwidth]{histograms}$$
%\caption{\em Distibutions of the ratios between the approximate over
%true  bound in various approximations. In the first {\it ``oblique''} panel
%we include in the fit only $\hat S$, $\hat T$,
%$\hat U$, $W$, $Y$, $V$ and $X$. In the second panel we add the two parameters
%$\delta C_q$ and $\delta \varepsilon_q$ for the quarks. Finally we include all the parameters except $\delta C_q$ and $\delta \varepsilon_q$.}
%\label{fig:histo}
%\end{center}
%\end{figure}

We now check how good our approximations are for guessing
the bound on the scale $\Lambda$ of new physics in generic models. To do
that, we generated many random models
by writing each coefficient $c_i$ of eq. (\ref{eq:lag}) as $c_i = r_i/\Lambda^2$, where
$-1\le r_i \le 1$ are random numbers.
We then extract the bound on $\Lambda$ both from the exact fit and the approximate
fits. In the
following table we report the average value and the variance of
$\Lambda_{\rm approx}/\Lambda_{\rm true}$ in the three following cases: the  oblique approximation, then we add the
two parameters $\delta C_q$ and $\delta \varepsilon_q$ for the
quarks and finally we include all the
parameters except $\delta C_q$ and $\delta \varepsilon_q$.

\vspace{0.3cm} \begin{center}
\begin{tabular}{c|c}
Approximation &  $\Lambda_{\rm approx}/\Lambda_{\rm true}$ \\
\hline
Oblique & $0.95 \pm 0.16$ \\
Oblique plus $C_q$, $\delta \varepsilon_q$   & $0.98 \pm 0.06$ \\
All but $C_q$, $\delta \varepsilon_q$  & $0.98 \pm 0.15$
\end{tabular}
\end{center} \vspace{0.3cm}
We see that the oblique approximation is already
reasonable:  in most of the cases the approximate bound is less than
$25 \%$ away from the correct one. Adding the two parameters $\delta
C_q, \delta \varepsilon_q$ improves the approximation significantly:
in more than $90 \%$ of the cases the approximate bound reproduces
the exact one within $10 \%$. Furthermore, it is important to notice
that considering a fit where all the parameters except $\delta C_q ,
\delta \varepsilon_q$ are added does not improve much the
approximation with respect to the oblique case. This is telling
us that in the quark sector it is indeed $\delta C_q$ and $ \delta
\varepsilon_q$ which are the most constrained parameters, while all
the others are much less constrained (and mostly negligible). The arguments of Sec. \ref{sec:parameters} which led us to define the 9 parameters find here a quantitative confirmation: the 9 remaining parameters can be safely
neglected.

\section{An example: $Z'$s}
\label{sec:Zprimes}

To be concrete, we apply our analysis to a generic heavy non-universal $Z'$ vector boson. It is characterized by its mass $M_{Z'}$, its coupling $g_{Z'}$ and charges $Z_X$ under the various SM fields $X = \{H, E, L, Q, U, D\}$. In terms of these parameters one can easily calculate the corrections to the 9 parameters $\hat{S},\hat{T},\hat{U},V,X,W,Y,\delta \varepsilon_q, \delta C_q$ defined above \footnote{For the analytical expressions see Sec. 4 of \cite{Cacciapaglia:2006pk}}. In order to verify the goodness of our approximation it is interesting to compare the bounds on the ratio $M_{Z'}/g_{Z'}$ in the three following cases: i) exact case (i.e. including all the 20 parameters), ii) oblique approximation (i.e. including only $\hat{S},\hat{T},\hat{U},V,X,W,Y$), and iii) adding $\delta \varepsilon_q, \delta C_q$ for the quarks. The results are shown in Table \ref{tab:famZprimes}. It is interesting to notice that the approximate
bounds reproduce the exact one accurately in almost all the cases.
There are few exceptions where the effect of quarks is relevant, and
the oblique bound is overestimated. On the other hand, the
9-parameter approximation is always successful.

\begin{table}[t]
$$
  \begin{array}{cc|cccccc|ccc}
\hbox{U(1)}& \hbox{universal?}& Z_H & Z_L & Z_D & Z_U & Z_Q & Z_E & \hbox{full} &\hbox{approx} & \hbox{oblique} \\ \hline
H & \hbox{yes} & 1 & 0 & 0 & 0 & 0 & 0 & 6.7 & 6.7 & 6.7\\ 
B' & \hbox{yes} & \frac{1}{2} & -\frac{1}{2} & \frac{1}{3} & -\frac{2}{3} & 
\frac{1}{6} & 1 & 6.7 & 6.7 & 6.7\\ 
B'_F & \hbox{yes} & 0 & -\frac{1}{2} & \frac{1}{3} & -\frac{2}{3} & 
\frac{1}{6} & 1 & 4.8 & 4.8 & 4.8\\ 
B-L & \hbox{no} & 0 & -1 & -\frac{1}{3} & -\frac{1}{3} & \frac{1}{3} & 1 & 
6.7 & 7.1 & 7.1\\ 
L & \hbox{no} & 0 & 1 & 0 & 0 & 0 & -1 & 6.3 & 7.1 & 7.1\\ 
10 & \hbox{no} & 0 & 0 & 0 & 1 & 1 & 1 & 2.5 & 2.9 & 3.4\\ 
5 & \hbox{no} & 0 & 1 & 1 & 0 & 0 & 0 & 3.8 & 3.2 & 5.6\\ 
Y & \hbox{no} & \frac{2}{3} & 1 & 1 & -\frac{1}{3} & -\frac{1}{3} & 
-\frac{1}{3} & 4.8 & 5.0 & 6.0\\ 
16 & \hbox{no} & 0 & 1 & 1 & 1 & 1 & 1 & 4.4 & 4.7 & 6.5\\ 
R & \hbox{no} & 0 & 0 & -\frac{1}{2} & \frac{1}{2} &0 & -\frac{1}{2} & 1.6 & 1.5 & 1.7\\ 
\hbox{SLH} & \hbox{no} &  \multicolumn{6}{c|}{\hbox{Simplest little Higgs~\cite{Schmaltz:2004de}}} & 2.7 & 
2.5 & 2.7\\ 
\hbox{SU6} & \hbox{no} &  \multicolumn{6}{c|}{\hbox{Super little Higgs~\cite{Csaki:2005fc}}} & 3.1 & 3.3 
& 3.3\\  \end{array}$$
  \caption{\em $99\%$\CL\ bounds on the ratio $M_{Z'}/g_{Z'}$ in {\rm TeV} for a set of frequently studied $Z'$. In the last two cases we report the bound on the scale $f$ of little-Higgs models.}
  \label{tab:famZprimes}
\end{table}

\section{Conclusions}
\label{sec:conclusions}

The astonishing (and boring) success of the Standard Model has put many constraints on models aiming to solve the Higgs hierarchy problem. Most of them are today still compatible with the data only in rather unnatural regions of their parameter space. We have defined a minimal and exhaustive set of parameters which are constrained by the electroweak data. We have shown that experiments below, at and above the $Z$ pole are sensitive (assuming flavor universality and CP conservation) to 20 dimension 6 operators. However, only about half of them are severely constrained. Using the equations of motion for the gauge bosons we have defined a set of $7+2+1=10$ parameters. The 7 oblique parameters $\hat{S},\hat{T},\hat{U},V,X,W,Y$ include all the effects on the purely leptonic observables. 2 additional parameters $\delta \varepsilon_q, \delta C_q$ include the effects on the two most precisely measured hadronic quantities: the $Z$ hadronic branching fraction and the $e \bar e \to$ hadrons total cross section. Furthermore, an additional parameter $\delta \varepsilon_b$ can be added, describing the deviations in the coupling of the left handed bottom quark to the $Z$, if flavor universality is not assumed for the third generation of quarks.

We have then checked how good the various approximation is in two different ways. First, we have generated many random models by varying randomly the coefficients of the various parameters. Then, we applied our formalism to a set of famous $Z'$s. The result is that the purely oblique approximation (i.e. the 7 parameters $\hat{S},\hat{T},\hat{U},V,X,W,Y$) is often satisfactory. Adding the two parameters $\delta \varepsilon_q, \delta C_q$ for the hadronic observables improves significantly the oblique approximation: in more then 90 \% of the randomly generated models the correct bound is reproduced within 10 \%.

\section*{Acknowledgments}

I warmly thank G. Cacciapaglia, C. Csaki and A. Strumia
for their precious collaboration. It is also a pleasure to thank the
Organizing Committee of the XLIst Rencontres de Moriond.

\end{document}